Study of the experimental and simulated vibrational spectra together with conformational analysis for thioether cyanobiphenyl-based liquid crystal dimers

*Barbara Loska,[1] Antoni Kocot,[1] Yuki Arakawa,[2] G.H. Mehl,[3] Katarzyna Merkel,[1]\**

[1] Institute of Materials Engineering, Faculty of Science and Technology, University of Silesia, ul. 75 Pułku Piechoty, Chorzów 41-500, Poland

[2] Department of Applied Chemistry and Life Science, Graduate School of Engineering, Toyohashi University of Technology, Toyohashi, 441-8580, Japan

[3] Department of Chemistry, University of Hull, Hull HU6 7RX, UK

Abstract

Infrared spectroscopy (IR) and quantum chemistry calculations that are based on the density functional theory (DFT) have been used to study the structure, molecular interactions of the nematic and twist-bend phases for thioether-linked dimers. Infrared absorbance measurements were conducted in a polarized beam for a homogeneously aligned sample in order to obtain more details about the orientation of the vibrational transition dipole moments. The distributions to investigate the structure and conformation of the molecule dihedral angle were calculated. The calculated spectrum was compared with the experimental infrared spectra and as a result, detailed vibrational assignments are reported.

1. Introduction

Vibrational spectroscopy has become one of the most informative tools in the study of liquid crystal materials [1-10]. An analysis of the intensity and position changes of the vibrational bands makes it possible to identify the liquid crystal phases and to calculate the orientational order parameters. Unfortunately, for large molecules, it is virtually impossible to reliably define the vibrational fundamentals without any theoretical input; therefore, determining the vibrational frequencies using computational methods is extremely important. These methods are helpful in interpreting the experimental vibrational spectra for large molecules. They also give information about the shape of the bands and the orientation of the transition dipole moment in the molecular coordinate system. In this research, we studied the liquid crystal dimers that are based on cyanobiphenyl mesogens. These dimers typically contain two rigid terminal groups that are chemically linked to each other by a flexible spacer with an odd number of methylene units [11-13]. Interest in liquid crystal dimers is high because of their extraordinary flexoelectric [14-16] and electro-optical properties [17-20] and their ability to form modulated nematic phases ($N_{TB}$ – twist-bend, $N_{SB}$ – splay-bend phase) [21-28]. The twist-bend modulated nematic phase ($N_{TB}$) has a helical structure with a pitch length of several nanometers [18,22, 29-31]. The structure of the $N_{TB}$ phase has primarily been studied using non-resonant (SAXS, WAXS) [32-34] and resonant X-ray scattering [29,30,35-38] as well as using polarized Raman [39], infrared [40-43] and nuclear magnetic resonance spectroscopy (NMR) [44-48]. Current experimental results show the formation of the $N_{TB}$ phase is very sensitive to changes in the shape of the molecules [49-54], therefore the key is to define carefully the relationship between the molecular parameters of the molecules and the occurrence of the modulate nematic phase. For this purpose, many authors have used molecular modeling methods, mainly the density functional theory (DFT) [25,26,30,31,49,55-58] or molecular dynamics simulations (MD) [59-61] to analyze structural and conformational changes of the dimers. However, no analysis of the conformational changes of molecules in the transition from the nematic to the twist-bend nematic phase based on a comparison of simulated and experimental vibrational spectra has been performed so far. Polarized measurements of the vibrational spectra (Raman and infrared) are extremely useful in studies of the supramolecular systems that primarily arise through intermolecular interactions such as hydrogen bonding [42,62-67] and π-π-stacking [68]. The analysis of the vibrational spectra for such supramolecular systems, the modulated nematic phases are clearly examples of that, in the first stage, requires the assignment of the bands in the spectrum with regard to the individual functional groups in the molecule.

Here, we report on a spectroscopic and theoretical study for two groups of dimers: a symmetric one that contains the thioether-linking groups (C-S-C) and the ether-linking groups (C-O-C) and an asymmetric one that contains both the ether- and thioether-linking groups. By using the polarized infrared absorbance method for a homogeneously aligned sample, it was possible to obtain information about the orientation of the transition dipole moments of a molecule. This information was compared with the theoretically calculated cartesian components of the vibrational transition dipole moment for specific vibrations and has proven to be very useful in the precise assignment of the bands in experimental spectra.

2. Materials and Methods

2.1. Materials

The symmetrical and asymmetrical liquid crystal dimers with the cyanobiphenyl (CB) mesogenic groups were investigated. We present three symmetric dimers, which had the general acronym CBXC7XCB (X=C or S or O), that contain nine functional groups in the chain linking the two mesogenic core (methylene, thioether, ether): the CBC9CB that contain nine methylene groups in the linker and the CBSC7SCB and CBOC7OCB in which the alkyl chains that contained seven methylene groups are connected to the cyanobiphenyls by two thioether or two ether bridges, respectively. In the asymmetric dimers with the acronym CBSCnOCB (n=5,7), the mesogens are linked to an alkyl chain on one side with five or seven methylene groups by a thioether bridge and on the other by an ether bridge. The details for the materials CBC9CB were reported earlier [69-71 Synthetic details concerning the thioether/ether compounds were recently [56,72]. Figure 1 shows the chemical structure of the investigated compounds.

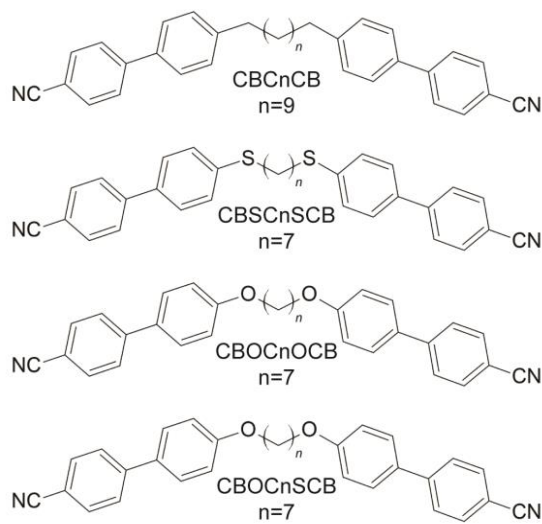

Fig.1. Chemical structures of the investigated dimers.

2.2. Infrared spectroscopy

The planarly aligned cells were prepared between two optically polished zinc selenide (ZnSe) discs. The thickness of the fabricated cells was determined to be within the range of 5.1-5.6 μm by the measurements on the interference fringes using a spectrometer that was interfaced with a PC (Avaspec-2048). The infrared spectra were acquired using a Fourier infrared spectrometer (Agilent Cary 670 FTIR). The experiment was conducted using the transmission method with a polarized IR beam. An IR-KRS5 grid polarizer was used to polarize the IR beam. The IR spectra were measured as a function of the polarizer rotation angle in the range 500-4000 cm$^{-1}$ of wavenumbers. The details of the sample preparation and absorbance measurement are published in Kocot et. Al [43]. These measurements enabled the orientation of the transition dipole moment of the bands to be determined with respect to the long molecular axis and the temperature dependencies of the absorbance of the samples. To determine all three components of absorbance ($A_x$, $A_y$, $A_z$), it is necessary to measure two samples with different orientations: planar (homogeneous) and homeotropic. Unfortunately, in the case of the tested materials, i.e., for the cyanobiphenyl dimers, it was extremely difficult to obtain a good homeotropic alignment. Therefore, in order to calculate the mean absorbance of the sample and assuming that the material was uniaxial, it was assumed that $A_x = A_y$ and therefore the mean absorbance was determined as $A_0=(2A_y+A_z)/3$. The absorbance components were determined as the area that was bound by the contour of a given band using Bio-Rad Win-IR Pro version 2.96e. In the case of complex bands that contained more vibrations, they were separated using Origin Pro 2021 software with the Pearson VII fit. Figure 1 shows the configuration of the infrared measurements using the polarized transmission technique (Fig.1a) and the molecular structure of the cyanobiphenyl dimers (Fig.1b).

2.3. Density functional theory calculations

In this work, the calculations of the electronic structure of the molecules were performed using the Gaussian09 program (version E.01) [73]. The molecular structures, intermolecular binding energy, harmonic vibrational force constants, absolute IR intensities and components of the transition dipole moment were calculated using the density functional theory (DFT) with the Becke's three-parameter exchange functional in combination with the Lee, Yang and Parr correlation functional B3-LYP method with the basis set: 6-311(d,p) [74]. The results were visualized using GaussView 5.0.8.

All of the DFT optimizations were performed with the following convergence criteria, which were used with the Berny algorithm (all values in atomic units): the maximum component of the force was set to 0.00045, the root-mean square (RMS) of the forces that were calculated for the next step - smaller then 0.0003; the computed displacement for the next step - smaller than 0.0018 and the RMS of the displacement below 0.0012. These criteria restricted the dependence of the final geometry parameters on the initial starting geometry.

2.3.1 *Molecular structures*

The rotational potential barriers of the molecules were obtained using the relaxed potential energy surface scan method with the molecular geometry optimized in order to find the most probable conformations that occurred in the tested materials. In order to find the most stable conformation of a dimer, the optimization of the geometry was performed in a few steps. All of the possible conformations of the dimers that were considered were defined by the values of the dihedral angles $\varphi_1$-$\varphi_4$ (Fig. 3b). In the first stage, the energy barriers for the internal rotation of the cyanobiphenyl (torsional angles $\varphi_3$ and $\varphi_4$) were determined. In a further step, the energy barriers for the rotation around the dihedral angle ($\varphi_1$, $\varphi_2$) between the cyanobiphenyl and the linker were determined. The approximate potential energy functions were calculated at intervals of ten. For the calculations, the torsional angles ($\varphi_1$-$\varphi_4$, each in turn) were fixed at arbitrarily selected values while the other geometrical parameters were optimized after which the relaxed potential energy scans were performed. This procedure enabled the values of the torsion angles for which the minimum energy was obtained to be determined. As the energy barrier for the internal rotation in the alkyl chain is very small (approx. 1 kJ/mol), no other linker/alkyl chains conformations than that the all-trans were considered. In the ordered phases, in principle, the alkyl chains can take upof the possible conformations, and therefore adopting all of the all-trans conformations are a better representation of the average molecular shape of the dimers [75]. In the next step, taking into account the values of torsion angles that were determined, a full optimization of the geometry was performed for all of the dimers.

*2.3.2. Molecular vibrations*

To perform an in-depth analysis of the experimental spectra, the density functional theory (DFT) was used to calculate the theoretical IR spectra for an isolated molecule. Generally, however, there are no absolute assignments of the IR frequencies, and for low symmetry molecules, the correlations of the calculated and experimental frequencies are primarily made by comparing bands that have similar frequencies, assuming that there is no rearrangement of the individual peaks in the bands.

All of the calculated vibrational frequencies are expressed as the wavenumber in cm$^{-1}$, while the so-called integral absorption coefficient, which is directly proportional to the sum of components of the squares of the transition dipole moments and this is represented by the formula:

$$\bar{A} = \frac{N\pi}{3c^2} \left[ \left(\frac{\partial \mu_x}{\partial Q}\right)^2 + \left(\frac{\partial \mu_y}{\partial Q}\right)^2 + \left(\frac{\partial \mu_z}{\partial Q}\right)^2 \right] \qquad (1)$$

where N – Avogadro constant, c – speed of light and $\frac{\partial \mu_i}{\partial Q}$ – the change of the component of the dipole moment with respect to the normal coordinates.

For a more precise analysis of the spectra, it was necessary to calculate the components of the transition dipole moment. Information about the components of the transition dipole moment for a specific vibration enables the parallel and perpendicular components of the spectral density to be calculated. The parallel component of the absorption coefficient was calculated as the square of the component of the transition dipole moment along the axis that coincided with the long axis of the dimer ($|\mu_z|^2$). To determine the perpendicular component of the spectral density, the sum of the squares of the transition dipole moments along the perpendicular directions was used ($|\mu_x|^2 + |\mu_y|^2$).

In order to compare the theoretical bands with the experimental results, the discreet spectrum lines were extended using the Gaussian function with a half-width of 7 cm$^{-1}$. Then, the frequencies were rescaled using the scaling coefficient that was determined for the base that was used, which is 0.967 ± 0.021, according to Computational Chemistry Comparison and Benchmark Database. The intensities of the bands were divided by the intensity of the highest band in the theoretical spectrum for a given molecule (or the conformation in the case of molecules for which two stable conformers were calculated). The results of the calculations were compared with the experimental data from the nematic phase in order to identify the most important bands in the spectrum.

3. Results and discussion

### 3.1. Dihedral angle distributions and conformations

Figure 2 shows the calculated rotational potential barriers. First, the torsion angle of the cyanobiphenyl was calculated ($\varphi_3, \varphi_4$). The calculated value of the angle for the minimum energy was 40° and the potential barriers to this twisted conformation were 8.71 kJ/mol for $\varphi = 0°$ and 8.98 kJ/mol (for $\varphi = 90°$). The value of the $\varphi_1/\varphi_2$ angle is the most crucial for determining the bend angle of the molecule. There are a number of conformations for the CBSCnSCB and CBSCnOCB dimers, which are important for calculating the orientational order, primarily due to possible dihedral angle ($\varphi_1, \varphi_2$): $C_{Al}$-S-$C_{Ar}$-$C_{Ar}$ of the sulfur bridge for CBSCnSCB and also $C_{Al}$-O-$C_{Ar}$-$C_{Ar}$ (*Ar* – aromatic ring, *Al* – alkyl) of the oxygen bridge for CBSCnOCB. The conformational energies of the individual molecules were calculated with respect to those dihedral angles. Both angles were found to have a minimum energy for the "zero" angle (planar conformation) in contrast to the corresponding angle for CBCnCB: $C_{Al}$-$C_{Al}$-$C_{Ar}$-$C_{Ar}$, which was close to 90° with a potential barrier of 5.5 kJ/mol (upright conformation). The energy barrier for the rotation of the $C_{Al}$-O-$C_{Ar}$-$C_{Ar}$ bridge was high enough (ΔU~13.7 kJ/mol) to enable only the planar conformer to be possible. In the case of the sulfur bridge ($C_{Al}$-S-$C_{Ar}$-$C_{Ar}$), however, the barrier was much smaller ΔU~2.1 kJ/mol), and therefore, the intermolecular interactions might have influenced the molecular structure, which is in agreement with Y. Cao et al. [31]. This means that for the dimers that contain sulfur more than one energetically stable conformation is possible. As the energy barrier for the internal rotation in the alkyl chain is very small (approx. 1 kJ/mol), no spacer other than the all-trans was considered. In the ordered phases, in principle, the alkyl chains can occupy all of the possible conformations, and therefore adopting all of the all-trans conformations are a better representation of the average molecular shape of the dimers [75].

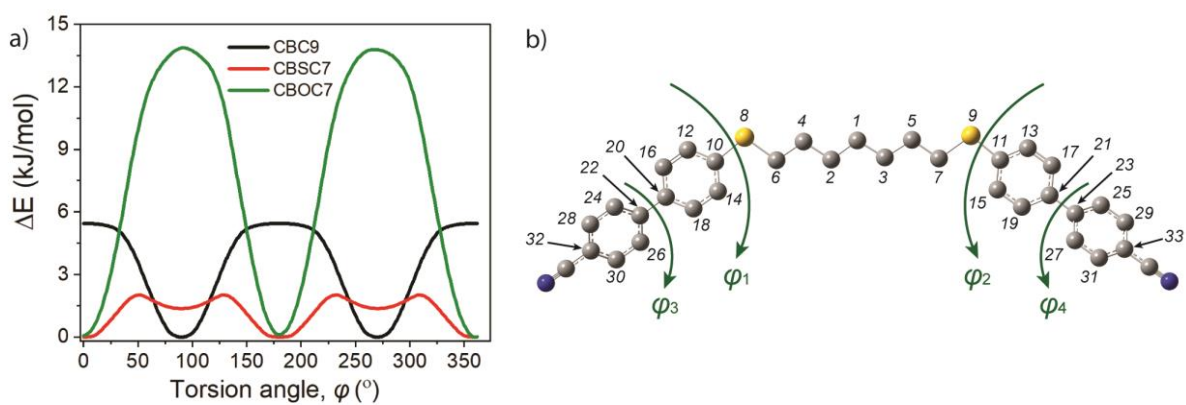

Figure 2. a) The calculated potential energy functions for the torsional motion of the cyanobiphenyl dimers relative to the minimum value ($\varphi_1$ or $\varphi_2$). *Black solid line* – CBC9CB dimer, *red solid line* – CBS7SCB dimer, *green solid line* – CBO7OCB dimer; b) Definition of the investigated dihedral angles of the dimers.

Taking into account the designated torsion angles, a full optimization was performed for each conformation of the dimers. Based on the coordinates of the atoms in a space, the coordinates and the lengths of the vectors corresponding to the arms of the molecules, the opening angles of the molecules were estimated for all of the probable conformations of the molecules (Fig. 3). Table 1 shows the torsion angle values and opening angle values for the fully optimized geometry of all of the dimers.

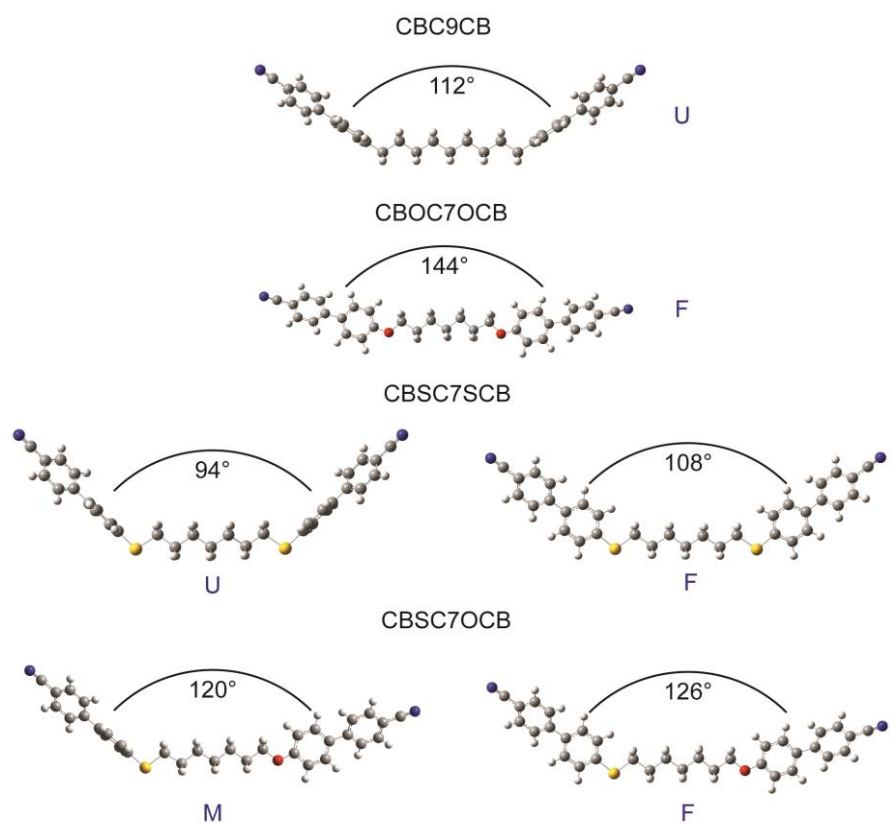

Figure 3. Probable conformers of the studied dimers. U – upright, F – flat (planar), M – mixed.

The bend of a molecule is necessary for the formation of the twist-bend phase. The largest opening angle (144°) was obtained for the CBOC7OCB molecule. This caused its shape to be closer to a calamitic molecule than to a bent (banana) molecule, which means that it did not meet the conditions for the formation of the twist-bend phase, which was confirmed experimentally *via* the POM, DSC and FTIR measurements [43,56,72,].

Table 1. Values of the most important torsion angles, potential energies and opening angles for the optimized geometry of a dimer.

| Sample | Conf. | Potential energy RMS (kJ/mol) | Kąty torsyjne, $\varphi_t$ (°) | | | | Opening angle (°) |
|---|---|---|---|---|---|---|---|
| | | | $\varphi_1$ | $\varphi_2$ | $\varphi_3$ | $\varphi_4$ | |
| **CBC9CB** | Upright (U) | -3843952.076 0.014 | 92.3 | 86.2 | -38.7 | -38.5 | 112 |
| **CBSC7SCB** | Upright (U) | -5521962.5193 0.0035 | -92.9 | -92.8 | -39.0 | -39.0 | 94 |
| | Flat (F) | -5521963.9252 0.0024 | 5.8 | 5.8 | -37.2 | -37.2 | 108 |
| **CBSC7OCB** | Mixed (M) | -4673985.9556 0.0077 | -96.1 | -179.1 | -38.8 | -37.2 | 120 |
| | Flat (F) | -4673986.6314 0.0037 | -176.3 | -180.0 | -37.4 | -37.6 | 126 |
| **CBOC7OCB** | Flat (F) | -3826009.436 0.020 | -179.3 | -179.2 | -37.3 | -37.3 | 144 |

### 3.2. Vibrational spectra and their assignments

It is practically impossible to obtain perfect order in an experiment, and additionally, the dichroism of the bands is influenced by many factors, mainly the intermolecular interactions. Therefore, based only on the analysis of the experimental spectra, it is difficult to say with absolute certainty which band and which of the cyanobiphenyl para axis will be characterized by the behavior of the long axis of a dimer, i.e., to independently describe the direction of the dimer arms. In the case of the theoretical spectra for symmetric dimers, for a given band of the cyanobiphenyl, two characteristic vibrations with a similar frequency but with different intensities should always be obtained. These correspond to the simultaneous vibrations in both arms, one of which is in the phase and the other is out of the phase. These vibrations, while coupling with each other, can ultimately cause a transition dipole moment to be along the long axis of the dimer or across it. Based on the analysis of the band dichroism for the theoretical spectra, which illustrate the theoretical, "ideal" order, it can be said that if the so-called infinite dichroism can be obtained, i.e., when one of the components is maximum and the other is close to zero, then such a band will then describe the behavior of the long axis of a dimer. On the other hand, in any other case, when the band dichroism is intermediate, there is a lack of vibration coupling for both arms and this band will describe the behavior of the para axis of the mesogens. The experimental spectra in the entire range of wavenumbers (500 – 3500 $cm^{-1}$) for all of the studied dimers in the nematic phase are presented in Fig. F1S in the *Supplementary materials*. Overall, the spectra of the cyanobiphenyl dimers can be divided into the following frequency ranges:

- 500-600 $cm^{-1}$ and 700-900 $cm^{-1}$ ranges, which covers the deformational vibrations of the carbon atoms (C-C) and hydrogen atoms (C-H) out of the benzene plane;
- 900-1650 $cm^{-1}$ range, which includes both the characteristic deformation vibrations in the benzene plane as well as the deformation vibrations of the methylene groups of the alkyl chain of the linker in a dimer;
- 2100-2400 $cm^{-1}$ range, which includes the stretching vibrations of the cyan group (C≡N) that were observed as a sharp and very intense peak in the experimental spectrum;

- 2800-2950 cm$^{-1}$ range, which includes the C-H stretching vibrations of the methylene groups. In this range, the vibrations were not well reproduced by the theoretical spectra because the calculations did not take into account the anharmonic effect.
- 2900-3100 cm$^{-1}$ range, which represents the stretching vibrations of the hydrogen atoms (C-H) in the aromatic ring. These vibrations were also not well reproduced by the theoretical frequencies. The bands in this range corresponded to the mixed vibrations, which were strongly overlapping, and the vibrations stretching the C-H hydrogen atoms, which were strongly disturbed by the Fermi resonance effect.

Most of the fundamentals in the range of 500 – 2300 cm$^{-1}$ were very well reproduced by the vibrations in the experimental spectra. Most of the observed changes relative to the transition from the nematic phase to the twist-bend and with the conformational changes that were observed for the thioether dimers in the range from 500 to 1650 cm$^{-1}$.

In this section, we will present the spectra analysis for the CBS7SCB and CBS7OCB dimers in more detail. A comparison of the experimental and theoretical spectra for the dimers, CBC9CB and CBOC7OCB, are summarized in the *Supplementary materials* (Fig 2S and Fig 3S). Table 2 presents a comparison of the main experimental bands and their assignments to the vibrations of the appropriate functional groups for all of the studied dimers.

*3.2.1.    The CBSC7SCB dimer*

Figure 4 shows the comparison between the polarized experimental and theoretical spectra for the CBSC7SCB dimer in the region of 500–1650 cm$^{-1}$. The most important differences in the theoretical spectra between the two conformations (flat and upright) were observed in the range of 500-1000 cm$^{-1}$. In the low frequency range, two medium-intensity bands at 520 and 560 cm$^{-1}$, which were assigned to the deformational vibration of the carbon atoms out of the benzene plane (*γCC op CB*; op – out of benzene plane CB-cyanobiphenyl) and a low-intensity band on the slope of the 560 band at the wavenumber of 551 cm$^{-1}$, which referred to the deformational vibration of the C≡N group (*δCN*) were observed. In the experimental spectra, the band at the wavenumber of 520 cm$^{-1}$ had a perpendicular direction of the transition dipole moment. This band also involved a thioether bridge and made a significant contribution to the deformation vibration of the sulfur atom (*γCC op CB* + *δCS*). In the theoretical spectrum for the planar conformation (F), as a result of the coupling of the vibrations of both arms, a band at the wavenumber of 514 cm$^{-1}$ was observed with the maximum dichroism and the transition dipole moment was perpendicular to the long axis of the dimer. In the case of the second conformation, which was called upright (U), two bands of comparable intensities were observed, one representing the vibration in the phase of both arms, which indicates the perpendicular direction of the transition dipole moment, while the antiphase vibration of both arms indicated the direction of the transition dipole moment parallel to the long axis dimer. Therefore, this band did not show dichroism for the U conformation. Additionally, for the upright conformation (U) in the theoretical spectrum, for the band at 650 cm$^{-1}$, we observed a splitting into two maxima, which was not observed in the experimental spectra.

In the range of 800 to 1120 cm$^{-1}$, the biggest differences between the conformers were associated with the out of plane deformation vibrations of the C-H groups (811 cm$^{-1}$; *γCH op CB*) and the deformation C-H vibrations in the benzene plane (1100 cm$^{-1}$). In the case of the upright conformation (U), there were two vibrations with similar wavenumbers and comparable intensities, which were similar to the 520 cm$^{-1}$ band. In the experimental spectra, the 811 cm$^{-1}$ band indicated the perpendicular direction of the transition dipole moment with a high dichroism, which corresponded to the flat conformer (F). The band at 1100 cm$^{-1}$ was a complexed vibration and involved a sulfide bridge and was assigned to the asymmetric stretching vibration of the C$_{Ar}$-S group (*βCH in CB* + *ν$_{as}$ C$_{Ar}$-S*; Ar – aromatic ring; as – asymmetric, in – in the benzene plane), therefore, this band was quite sensitive to the dimer conformational changes. Another indication that a flat conformation was more favored was the absence of a band at 1100 cm$^{-1}$ for the U conformer. In this range, there was also a band of low intensity at 1000 cm$^{-1}$, which was attributed to the breathing deformation vibration of the carbon atoms in the benzene ring (*βCC*).

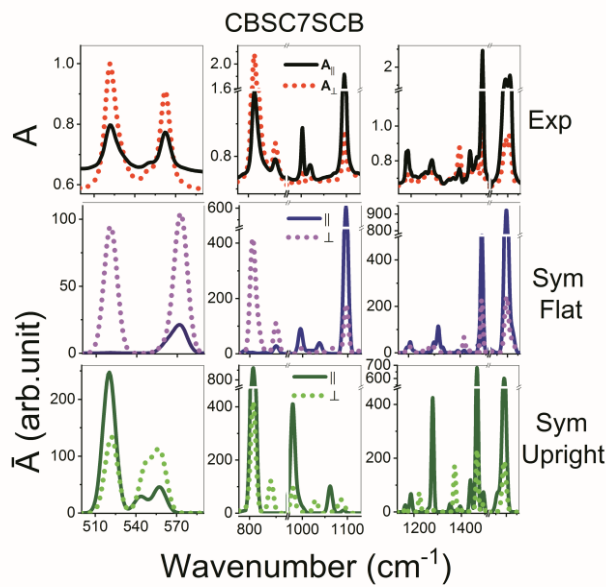

Figure 4. The comparison of the polarized experimental spectrum with the theoretical spectra (B3-LYP/6-311G (d,p)) for the CBSC7SCB dimer in the region of 500–1650 cm$^{-1}$. $A_{\parallel}$ – the parallel absorbance component in the z-axis direction of the molecular system, which coincided with the ordering axis of the sample. $A_{\perp}$ – the perpendicular absorbance component was perpendicular to the rubbing direction. *Top Figures* – experimental spectra of the nematic phase (370K). *Middle Figures* – calculated spectra for a planar conformer (dihedral angles $C_{Al}$-S-$C_{Ar}$-$C_{Ar}$ = 0°). *Bottom Figures* – calculated spectra for an upright conformation (dihedral angles $C_{Al}$-S-$C_{Ar}$-$C_{Ar}$ = 90°).

The next spectrum range that is discussed is the 1200-1650 cm$^{-1}$, where several high-intensity bands (1460, 1594, 1600 cm$^{-1}$) were observed, which were assigned to the benzene ring deformations ($\nu CC\ br$; br – benzene ring). They corresponded to the longitudinal transition dipole, which was similar to the bands at 1000, 1100 and 2300 cm$^{-1}$ wavenumbers (2300cm$^{-1}$ – assigned to the stretching vibrations of the cyan group, $\nu CN$). In this range, a low-intensity band with a wavenumber of about 1395 cm$^{-1}$ was also observed, the direction of which was the dipole transition perpendicular to the long axis of the dimer. This band was assigned to the in-plane deformation of the C-H groups ($\beta CH\ ip\ CB$). Differences between the two conformations for the benzene ring vibrations were also observed at 1600 cm$^{-1}$. In the experimental spectrum, this peak was split into two maxima with similar wavenumbers (1594 and 1604 cm$^{-1}$), for which the intensity shares changed with temperature. In the nematic phase, both peaks had a similar intensity, while in the twist-bend phase, an increase in peak intensity was observed at a lower wavenumber (1594 cm$^{-1}$). For the F conformer, this vibration was observed at a lower wavenumber, i.e., about 1595 cm$^{-1}$, while for the U conformer, it was observed at 1598 cm$^{-1}$ wavenumber. The above observations could lead to the conclusion that in the high temperature phases (nematic phase), the coexistence of both conformations and/or some intermediate conformers are probable, while when the temperature is lower, the planar conformation predominates, which is more preferred energetically.

### 3.2.2. *The CBSC7OCB dimer*

Figure 6 shows the comparison between the polarized experimental and theoretical spectra for the asymmetric CBSC7OCB dimer in the region of 500–1650 cm$^{-1}$. In the low frequency range, three medium-intensity bands were observed at 520, 530 and 560 cm$^{-1}$. It was observed that the band at the wavenumber 520 cm$^{-1}$, which characterized the vibration of the C-C groups out of plane together with the deformation vibration of the C-S group, had a perpendicular direction of the transition dipole moment in the experimental spectrum and was split into two maxima. An additional maximum appeared at the wavenumber 530 cm$^{-1}$, which was not observed for the CBSC7SCB dimer. This new band was assigned to the deformation vibration of the C-C group out of the benzene plane along with the deformation vibration of the C$_{Ar}$-O-C$_{Al}$ group *($\gamma$CC op CB + $\delta$C$_{Ar}$OC$_{Al}$*; Al – alkyl chain). In the theoretical spectrum of the F conformer, a split band was also obtained and the transition dipole moment was directed perpendicular to the long axis of the dimer. In contrast of the mixed conformation (M), only one undivided band was observed. Additionally, in the experimental spectrum, three bands were observed, which also corresponded to the transversal transition dipole moment: wavenumbers 813, 821 and 850 cm$^{-1}$. Compared to the CBS7SCB dimer, a new maximum appeared at 821 cm$^{-1}$, which was a complex band and in addition to involving the out of plane deformation vibrations of the C-H group, also described the symmetrical stretching vibration of the $C_{Ar}$-O group ($\gamma CH\ op\ CB + \nu_s\ C_{Ar}O$; s – symmetric). In the spectrum of the CBSC7OCB dimer, other new bands that were characteristic for the oxygen bridge were observed: asymmetric stretching vibrations of the $C_{Al}$-O group: ($\nu_{as}\ C_{Al}\ O$) at 1029 and 1050 cm$^{-1}$ and strong vibrations that were assigned to the asymmetric stretching of the C-O group ($\nu_{as}\ C_{Ar}O$) at 1249 and 1266 cm$^{-1}$.

In the 800-1120 cm$^{-1}$ range, the greatest differences between the conformers were associated with the out of plane vibration of the C-H group (813, 821 cm$^{-1}$). In the case of the mixed conformation, only one maximum of high intensity at 816 cm$^{-1}$ was observed. In the case of the F conformer, two maxima were observed, which was in good agreement with the experiment. Another indicator that a planar conformation is more likely is the absence of a maximum at 1100 cm$^{-1}$ in the spectra of M conformer, which was the same as for symmetric the dimer.

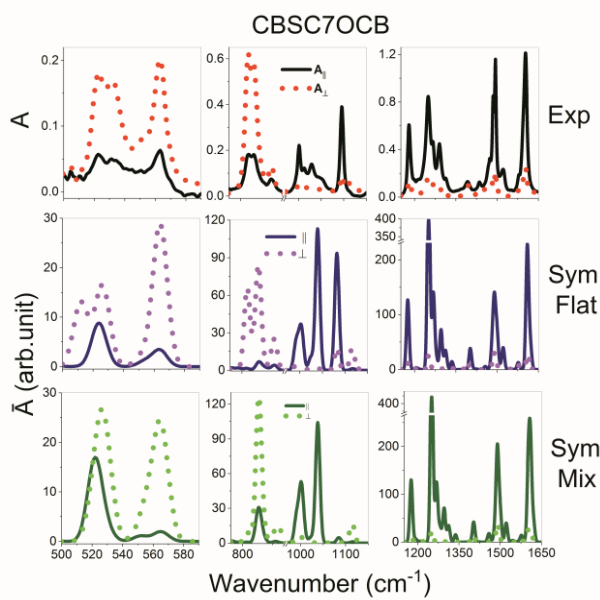

Figure 5. The comparison of the polarized experimental spectrum with the theoretical spectra (B3-LYP/6-311G (d,p)) for CBSC7OCB dimer in the region of 500–1650 cm$^{-1}$. $A_\parallel$ – the parallel absorbance component in the z-axis direction of the molecular system, which coincided with the ordering axis of the sample. $A_\perp$ – the perpendicular absorbance component was perpendicular to the rubbing direction. *Top Figures* − experimental spectra of the nematic phase (370K). *Middle Figures* – calculated spectra for a planar conformer (dihedral angles $C_{Al}$-S-$C_{Ar}$-$C_{Ar}$ = 0°). *Bottom Figures* – calculated spectra for an upright conformation (dihedral angles $C_{Al}$-S-$C_{Ar}$-$C_{Ar}$ = 90°).

Table 2. Vibrational frequencies, IR intensities and assignments of the dimers.

| CBC9CB | | CBSC7SCB | | CBSC7OCB | | CBOC7OCB | | ASSIGNMENTS |
|---|---|---|---|---|---|---|---|---|
| $\bar{\nu}$ | $I_r$ | $\bar{\nu}$ | $I_r$ | $\bar{\nu}$ | $I_r$ | $\bar{\nu}$ | $I_r$ | |
| 520 | vvw | 522 | m | 523 | w | -- | -- | $\gamma$CC op CB + $\delta$CS |
| -- | -- | -- | -- | 532 | w | 532 | w/m | $\gamma$CC op CB + $\delta$CO |
| 554 | w | 562 | w | 562 | w | 562 | w | $\gamma$CC op CB + $\delta$CN |
| 816 | vs | 811 | vs | 813 | vs | -- | | $\gamma$CH op CB |
| -- | -- | -- | -- | 821 | s,sh | 821 | vs | $\nu_s$COC + $\gamma$CH op CB |
| 836<br>850 | w | 852 | w | 850 | w | 850 | w | $\gamma$CH op CB + $\nu$CCC sk. + $\delta_{as}$CH$_2$ rocking |
| 1007 | w | 999 | w | 999 | w | 1000 | w | $\beta$CC ip CB, breathable |
| 1026 | vw | 1020 | vw | 1013 | vw | 1013 | vw | $\nu$CCC sk. + $\beta$CH ip CB |
| -- | -- | -- | -- | 1029<br>1051 | w | 1032 | w | $\nu_{as}C_{Al}$O + $\beta$CH ip CB |
| -- | -- | 1097 | m | 1095 | m | -- | -- | $\nu_{as}C_{Ar}$S + $\beta$CH ip CB |
| 1112 | vw | -- | -- | -- | -- | 1120 | w | $\beta$CH ip CB |
| 1185 | m | 1185 | s | 1180 | s | 1178 | s | $\beta$CH ip CB |
| -- | -- | -- | -- | 1249<br>1266 | vs<br>m,sh | 1249<br>1266 | vs<br>m,sh | $\nu_{as}C_{Ar}$O + $\beta$CH ip CB |
| 1284<br>1315<br>1360 | vw | 1279<br>1315 | vw | 1290<br>1311 | vw | 1290<br>1313 | vw | $\gamma_s$CH$_2$ wagging<br>$\gamma_s$CH$_2$ twisting |
| 1397 | w | 1395 | w | 1392 | w | 1390 | w | $\beta$CH ip CB |
| 1460 | w | 1437<br>1462 | w | 1435<br>1472 | w | 1472 | w | $\beta_s$CH$_2$ scissoring |
| --<br>1493 | s | 1484<br>-- | vs | 1485<br>1494 | vs | --<br>1493 | vs | $\nu$CC br |
| -- | -- | -- | -- | 1522<br>1577 | m<br>vw | 1523<br>1580 | m<br>w | $\nu$CC br + $\beta_s$CH$_2$ + $\nu_{as}C_{Ar}$O |
| --<br>1605 | s | 1594<br>1604 | vs | --<br>1603 | vs | --<br>1602 | vs | $\nu$CC br |
| 2224 | vs | 2223 | vs | 2223 | vs | 2223 | vs | $\nu$CN |

Key: ip – in plane vibration; op – out of plane deformation, br – stretching and deformation vibrations of the ring (benzene ring), s – symmetrical, as – asymmetric, Al. – alkyl chain, Ar – aromatic ring, $\nu$ – stretching, $\gamma$ – deforming out of plane, $\beta$ – deforming in plane, $\delta$ – deforming, vs – very strong, s – strong, m – medium, w – weak, vw – very weak, sh – shoulder. $I_r$ – relative intensity of the bands

A more detailed and row data of the theoretical and experimental frequencies, dichroism values, relative intensity, direction of the transition dipole moment and approximate band assignments for all of the investigated dimers are included in the *Supplementary materials* (Table 1S- 4S).

4. Conclusions

Using the DFT simulations of the electron structure of molecules, the rotational barriers were determined and the optimization of the most probable conformers that could appear in real samples was performed based on them. Two energetically stable conformations were found for the molecules that contained sulfide bridges and one was found for the molecules CBC9CB and CBOC7OCB. The opening angles of the tested dimers were also determined. The simulated theoretical spectra turned out to be an invaluable tool for identifying the most important bands in the experimental spectra as well as in determining the directions of the dipole transition moments of the molecules. Combined with the observations of the deflections in the normal modes of the atoms from the equilibrium positions, bands were assigned to the appropriate fragments of the molecules and thus information about the geometry of the system was obtained.

**AUTHOR INFORMATION**

Corresponding Authors: Katarzyna Merkel
* E-mail: katarzyna.merkel@us.edu.pl
E-mails co-authors:
Antoni Kocot: antoni.kocot@us.edu.pl
Barbara Loska: barbara.loska@us.edu.pl
Yuki Arakawa: arakawa.yuki.xl@tut.jp
Georg Mehl: g.h.mehl@hull.ac.uk

**ORCID:**

Antoni Kocot: 0000-0002-9205-449X
Barbara Loska: 0000-0002-0756-5018
Yuki Arakawa: 0000-0002-8944-602X
Georg Mehl: 0000-0002-2421-4869
Katarzyna Merkel: 0000-0002-1853-0996

**Author Contributions:**

Conceptualization: K.M
 Synthesis:  G.H.M, Y.A
Methodology: K.M, A.K
Investigation: K.M, B.L

Writing—original draft: K.M
Writing—review & editing: K.M, A.K

**Funding Sources:**

National Science Centre, Poland for Grant No. 2018/31/B/ST3/03609

**Competing interests:**

Authors declare that they have no competing interests.